\documentclass[12pt]{article}
\usepackage{amsmath,amssymb}
\usepackage{latexsym}

\begin{document}


\begin{center}
\LARGE \textbf{\emph{Understanding Permutation Symmetry.}}

\vspace{20pt}

\small STEVEN FRENCH \& DEAN RICKLES
\end{center}

\vspace{10pt} \hrule \vspace{30pt}


\begin{flushright}
\begin{quotation}
If a system in atomic physics contains a number of particles of
the same kind, e.g. a number of electrons, the particles are
absolutely indistinguishable one from another. No observable
change is made when two of them are interchanged ... A
satisfactory theory ought, of course, to count any two
observationally indistinguishable states as the same state and to
deny that any transition does occur when two similar particles
exchange places. (Dirac 1958, 207.)
\end{quotation}
\end{flushright}


\begin{center}
\textbf{1. Introduction.}
\end{center}

\noindent In our contribution to this volume we deal with
\emph{discrete} symmetries: these are symmetries based upon groups
with a discrete set of elements (generally a set of elements that
can be enumerated by the positive integers). In physics we find
that discrete symmetries frequently arise as `internal',
non-spacetime symmetries. Permutation symmetry is such a discrete
symmetry arising as the mathematical basis underlying the
statistical behaviour of ensembles of certain types of
indistinguishable quantum particle (e.g., fermions and bosons).
Roughly speaking, if such an ensemble is invariant under a
permutation of its constituent particles (i.e., permutation
symmetric) then one doesn't `count' those permutations which
merely `exchange' indistinguishable particles; rather, the
exchanged state is identified with the original state.

This principle of invariance is generally called the
`indistinguishability postulate' [IP], but we prefer to use the
term `permutation invariance' [PI]. It is this symmetry principle
that is typically taken to underpin and explain the nature of
(fermionic and bosonic) quantum statistics (although, as we shall
see, this characterisation is not uncontentious), and it is this
principle that has important consequences regarding the
metaphysics of identity and individuality for particles exhibiting
such statistical behaviour.

In this paper we will largely be dealing with the following two
types of problem:
 \vspace{5pt}

(1) \textbf{How are we to understand the metaphysics of PI}?

\vspace{10pt}

\noindent For instance, do we follow the `received view' and say
that permutation invariance shows us that quantum particles are
not individuals? Do we maintain that they are individuated by
their spatiotemporal location, or perhaps by some
extra-theoretical property (e.g., the `primitive thisness' of the
object)? Given this individuation how are we to understand PI?
Maybe we can resolve the issue in some completely different way,
with `structures' replacing `objects' perhaps? It is clear that
such questions readily relate to `traditional' metaphysical issues
connected to identity and individuality.

\vspace{10pt}

(2) \textbf{How are we to understand the status of PI,
theoretically and \\ \indent \;\;\;\; empirically}?

\vspace{10pt}

\noindent For example, should PI be considered as an axiom of
quantum mechanics? Or should it be taken as justified empirically?
Why do there appear to be only bosons and fermions in the world
when PI allows the possibility of many more types? This is usually
resolved by postulating, ad hoc, some `superselection rule',
called the ``symmetrisation postulate'' [SP], restricting the
state vector to the fermionic and bosonic subspaces of the
systems' Hilbert space. However, rather than resolving the
difficulty, this simply moves the explanatory task one step
backwards (i.e., how are we then to understand SP?).
Alternatively, the extra, possibly redundant, mathematical
structure responsible for the extra possibilities regarding
symmetry types of particles can be understood as `surplus
structure' (in the sense of Redhead 1975). One often finds such
surplus structure in theories possessing lots of symmetry, and it
frequently points to the existence of `gauge freedom' in a theory
(e.g., in general relativity, Yang-Mills theory, and
electromagnetism).  It is here that a possible relation of
permutation invariance to diffeomorphism invariance (the symmetry
underlying the general covariance of general relativity) becomes
apparent.

\vspace{5pt}

In this paper we survey a number of these issues and their
consequences, introducing the reader to the various schools of
thought regarding the status and interpretation of PI (and ,
likewise, though to a lesser extent, SP). Let us begin with a
brief introduction to the formal aspects of PI and relevant
related topics in group theory and classical/quantum statistical
mechanics.

\newpage
\begin{center}
\textbf{2. The Mathematics \& Physics  of Permutation Symmetry.}
\end{center}

Permutation symmetry is a discrete symmetry supported by the
permutation group Perm($\mathcal{X}$) of bijective maps (the
permutation operators, $\hat{P}$) of a set $\mathcal{X}$ onto
itself.\footnote{The fact that the set Perm$(\mathcal{X})$ has the
structure of a group simply means that: (1) we can combine any two
elements ($\hat{P}_{1} , \hat{P}_{2} \in$ Perm($\mathcal{X}$)) in
the set to produce another element ($\hat{P}_{3} = \hat{P}_{1}
\cdot \hat{P}_{2}$) that is also contained within that set
($\hat{P}_{3} \in$ Perm($\mathcal{X}$)); and (2) each element
$\hat{P} \in$ Perm($\mathcal{X}$) also has an inverse
$\hat{P}^{-1} \in$ Perm($\mathcal{X}$).} When $\mathcal{X}$ is of
finite dimension Perm($\mathcal{X}$) is known as the symmetric
group $S_{n}$ (where the $n$ refers to the dimension of the
group). For instance, $\mathcal{X}$ might be the set consisting of
the labels of the two sides of a coin: heads `H' and tails `T'. Or
perhaps the `names' of $n$ particles making up some quantum
mechanical system, an He$^{4}$ atom for example. If we take the
coin as our example, then $\mathcal{X} = \{H,T\}$ and
Perm($\mathcal{X}$) is an order two group, $S_{2}$, consisting of
two elements (computed as having 2! elements via the dimension, $n
= 2$, of the group): (1) the identity map, $id_{\mathcal{X}}$,
which maps H to H and T to T; and (2) the `flip' map (or
`exchange' operator), $\hat{P}_{HT}$, which maps H to T and T to
H.

Now, to say that some object (i.e. a set or the total state vector
of a system of particles) is `permutation symmetric' means that it
is invariant under the action of Perm($\mathcal{X}$): it remains
unchanged (in some relevant sense) when it is operated upon by the
elements (i.e., the permutation operators) of Perm($\mathcal{X}$),
including (for $n \geq 2$) the elements that `exchange' the
components of the object (in this case the labels of the sides of
the coin or the labels of the particles in a quantum system).

The coin clearly is not permutation symmetric (i.e., does not
satisfy PI), since we must distinguish `heads' from `tails'; that
is, there is an \emph{observable} difference between these two
states of a coin. However, when we consider systems containing
several indistinguishable particles\footnote{Particles are said to
be indistinguishable in that they possess the same state
independent (intrinsic) properties, such as rest mass, charge, and
spin. Since these quantities have a continuous spectrum in
classical mechanics we can still individuate particles by their
variations with respect to these properties. If it were the case
that we had a classical system containing particles that exactly
matched in these properties, then we could still distinguish the
subsystems by their spatiotemporal location. Such luxuries are not
available in quantum theory because of discrete spectra and the
absence of definite trajectories.}, each with several possible
states (particles such as electrons, neutrons, and photons), we
find that they are indeed permutation symmetric, and that this
symmetry `shrinks' the number of possible states of the total
system, thus altering the statistical behaviour of the ensemble.
In this way PI is generally taken to \emph{explain} the divergence
of quantum statistics from classical statistics.\footnote{Note,
however, that this explanatory link has been contested by Huggett
(1999a).}

To see how these `altered statistics' follow from PI, and what
they look like, let us compare classical and (bosonic/fermionic)
quantum statistics using a simple example.

Consider the distribution of a system of $n$ indistinguishable
objects (e.g., free particles) over $m$ microstates. It is helpful
at this stage to view the objects as balls and the microstates as
the two halves of a box (making each side big enough to
accommodate all $n$ balls).\footnote{The separators in the
diagrams are there as an aid to visualization rather than as a
part of the system we are considering.} Statistical mechanics is,
very loosely, the study of the number of ways one can redistribute
the objects over the microstates without altering the macrostate.
Let us consider the simple case where we have two objects (balls)
and two microstates (boxes). Let us label the balls by `$a$' and
`$b$', and the sides of the box by `L' (left) and `R' (right). Let
`L(a)' be the state where ball a is in the left hand side (LHS) of
the box; let `L(ab)' be the state where both balls are in the LHS;
and let `L(0)' mean that the LHS is empty (similarly,
\emph{mutatis mutandis}, for the right hand side (RHS) and ball
b). Classically, we have four possible distributions:

\begin{center}
\vspace{10pt}

\begin{tabular}{|c|c|}
  \hline
  L(a) & R(b) \\
  \hline
  L(b) & R(a) \\ \hline
  L(ab) & R(0) \\ \hline
  L(0) & R(ab) \\
  \hline
\end{tabular}

\vspace{10pt}
\end{center}

\noindent Each possible permutation of the balls is counted in the
statistics and, if we assume equiprobability, each configuration
has a probability of 1/4 of being realized. Such a distribution is
known as a Maxwell-Boltzmann distribution, and it follows the
corresponding statistics for such distributions.\footnote{Huggett
(op. cit.) has argued that Maxwell-Boltzmann statistics do not
necessarily imply that we must must count permutations as
distinct: when there are many states available to each particle
the rule breaks down. In the case of the present example we are
dealing with many particles per state, and so the relation between
Maxwell-Boltzmann statistics and counting permutations as distinct
still holds.}

    The situation is different when we consider quantum particles
    because, in addition to being indistinguishable, they are subject
    to PI. There are two types of statistical behaviour for particles
    in quantum mechanics having to do with the ways in which they can
    combine in ensembles.\footnote{Of
    course, we are, for the moment, ignoring the case of
    para-statistics; namely, types of quantum statistics that violate SP,
    on which see \S 3 and \S 4.}
    Firstly, we
    have bosons (particles with integer spins; e.g., photons) behaving
    according to the Bose-Einstein statistics: meaning,
    \emph{inter alia}, that these particles can occupy the same
    state in a quantum system (the balls can reside the same side of the box).
    Secondly, we have fermions
    (particles with half-integer spins; e.g., electrons) behaving
    according to the Fermi-Dirac statistics: meaning,
    \emph{inter alia}, that these particles \emph{cannot} occupy
    the same state in a quantum system (the balls cannot reside in
    the same side of the box). This latter principle - not directly
    connected to PI - is
    generally known as Pauli's Exclusion Principle.

    These two
    points have an impact on the possible configurations we can count in
    the statistics. For instance, in the case of bosons we
    identify those configurations which differ only by an exchange
    of identical particles (i.e., the first and second configurations from
    the classical statistics
    above), but we can allow those configurations in which two
    objects occupy the same state. So if we consider the balls as
    bosons we get the following three distinct possibilities
    (where `L(1)' means that `some' particle is in the LHS -
    similarly, \emph{mutatis mutandis} for the RHS):

\begin{center}
\vspace{10pt}

\begin{tabular}{|c|c|}
  \hline
  L(1) & R(1) \\
  \hline
  L(ab) & R(0) \\ \hline
  L(0) & R(ab) \\
  \hline
\end{tabular}

\vspace{10pt}
\end{center}

\noindent So we have removed a classically possible state by
identifying `exchanged states'.\footnote{Note that we have
simplified the first configuration here, since what we actually
have, formally, is the state: [(L(a) \& R(b)) + (L(b) \& R(a))].
In the fermionic case we find a similar superposition only with a
change in sign (when the permutations are odd): [(L(a) \& R(b)) -
(L(b) \& R(a))]. Note that the change of sign has no effect on the
observable properties (expectation values) of the system.} This
has the consequence that the probabilities for finding a system in
a certain state (still assuming equiprobability) each go from 1/4
to 1/3. Following a similar procedure with fermions, and then
applying the exclusion principle, we get just one possible state:

\begin{center}
\vspace{10pt}

\begin{tabular}{|c|c|}
  \hline
  L(1) & R(1) \\
  \hline
\end{tabular}

\vspace{10pt}
\end{center}

\noindent Which, of course, has a probability of 1 of being
realized. All we have done here is to identify those
configurations which differ only in which ball occupies which side
of the box (following Dirac's intuition expressed in the opening
quote) and then we have forbidden two balls to occupy the same
side. In both the quantum cases the systems (or, more formally,
their state vectors) are invariant under the action of the
permutation group: when we apply the permutation operators to the
state vectors they continue to describe the same physical state;
following Dirac's intuition we identify the states. Hence, the
quantum systems satisfy PI, unlike the classical system. Let us
now make some of these ideas more exact by introducing some
elementary quantum theory.

States of quantum systems (single or many-particle) are
represented by rays $\Psi$ in a Hilbert space $\mathcal{H}$. For
many particle systems the Hilbert space is the joint space
constructed by tensoring together the component particles' Hilbert
spaces. The observables $\hat{O}$ of a quantum system are
represented by Hermitian operators acting upon that system's
Hilbert space.

    Now consider a system consisting of two indistinguishable
    particles. The Hilbert space for this system is:
    $\mathcal{H}_{total} = \mathcal{H}_{1} \otimes \mathcal{H}_{2}$,
    where
    the subscripts
    `1' and `2' label the composite particles, and $\mathcal{H}_{1} =
    \mathcal{H}_{2}= \mathcal{H}$. If the particles are in the
    pure states
    $\phi$ and $\psi$ respectively, then the composite system is
    in the (pure)
    state $\Psi = \phi \otimes \psi$. The permutation operators
    act upon
    $\Psi$ as follows: (1) $\hat{P}_{id}(\Psi) = (\phi \otimes \psi)$
    and (2)
    $\hat{P}_{\phi \psi}(\Psi) = (\psi \otimes \phi)$.

      The Hamiltonian, $\hat{H}_{\Psi} =
    \hat{H} (\phi \otimes \psi)$, of the composite system is
    symmetric with respect to $\phi$ and $\psi$. Hence,
    $\hat{H}_{\Psi}$ is invariant under the action of the
    permutation group of permutations of
    the composite particles' labels:  $[\hat{H},\hat{P}] = 0$,
    $ \forall \hat{P}$.  By an invariance of a quantum state under
    the action of the permutation group (i.e., PI) we then
    mean that every physical observable $\hat{O}$ commutes
    with every permutation operator $\hat{P}$:
    $[\hat{O},\hat{P}] = 0$, $\forall \hat{O} \forall \hat{P}$ -
    the
    physical interpretation of this being that there is no
    measurement that we could perform which would result in a
    discernible difference between permuted (final) and unpermuted
    (initial) states.
    This has the consequence that expectation values for
    unpermuted states are equal to expectation values
    for permutations of that state. Or, more formally, for
    any arbitrary state $\psi$, Hermitian operator $\hat{O}$,
    and permutation operator $\hat{P}$:

\begin{equation}
\langle\psi\mid\hat{O}\mid\psi\rangle =
\langle\hat{P}\psi\mid\hat{O}\mid\hat{P}\psi\rangle =
\langle\psi\mid\hat{P}^{-1}\hat{O}\hat{P}\mid\psi\rangle
\end{equation}

\noindent It is this result - basically a formal expression of PI
- which motivates the claim that PI can be understood as a
restriction on the possible observables of a system given its
state and, as such, it can be viewed as a superselection rule
determining which observables are physically relevant. We shall
return to this claim and, more generally, the status of PI in
later sections.

Finally, let us turn to the mathematical representation of
particle types. For this we need the concept of an `irreducible
representation'. Firstly, a representation $\rho$ of a group $G$
on a linear space $V$ is simply a map that assigns to each element
of the group $g \in G$ a linear operator $\hat{O} (V)$ on the
space. When the linear space is the (joint) Hilbert space
$\mathcal{H}$ spanned by the states, $\{ \phi \otimes \phi , \phi
\otimes \psi , \psi \otimes \phi , \psi \otimes \psi \}$, of two
indistinguishable particles, and the group is the permutation
group, the representation will associate a unitary operator acting
on $\mathcal{H}$ (i.e., on the state vector $\Psi \in
\mathcal{H}$) to each permutation operator $\hat{P} \in$
Perm$(\mathcal{X})$. We can represent this schematically as
follows (beginning with the group element, then the representation
of that element, and finally the physical operation)\footnote{We
should point out that this way of doing things is an
oversimplification in the following respects: firstly, since the
joint states lie in a four dimensional Hilbert space they are
represented by 4-vectors, but here we are assuming that they are
2-vectors. Also, the permutation operators should properly be $4
\times 4$ matrices, here we write them as $2 \times 2$ matrices.
However, since nothing of import depends on this, we prefer to
keep things simple in this way to facilitate understanding.}:

\vspace{5pt}
(1) $ \hat{P}_{\phi \psi} \Psi \Rightarrow \left(%
\begin{array}{cc}
  0 & 1 \\
  1 & 0 \\
\end{array}%
\right) \Psi$ $\Rightarrow $ `exchanging the particles'.\\
\vspace{5pt}

(2) $ \hat{P}_{id} \Psi \Rightarrow \left(%
\begin{array}{cc}
  1 & 0 \\
  0 & 1 \\
\end{array}%
\right) \Psi$ $\Rightarrow $ `leaving them alone'. \vspace{5pt}

\noindent This pair of matrices gives a unitary representation of
Perm$(\mathcal{X})$ on $\mathcal{H}$. The only matrix that
commutes with both of them is the unit matrix or some scalar
multiple of the unit matrix. Representations of this kind are said
to be `irreducible'. Alternatively, a representation is said to be
irreducible if the only invariant subspaces it possesses are $\{ 0
\}$ and $\mathcal{H}$ (i.e., the zero vector and the whole space)
- where a subspace $\mathcal{H}^{\prime}$ of $\mathcal{H}$ is
invariant if $\Psi \in \mathcal{H}^{\prime}$ implies
$\rho(\hat{P}) \Psi \in \mathcal{H}^{\prime}$, $\forall \hat{P}$.

 We are interested in the irreducible representations of the
 permutation group because each such representation is
 `carried' by an irreducible subspace of the Hilbert space,
 where each such subspace is
invariant under the action of the permutation operators. Thus, the
subspaces represent symmetry sectors corresponding to the possible
types of permutation symmetry possessed by the particles whose
state vectors lie in that subspace. In the case we are considering
we find that the total Hilbert space is partitioned into two
subspaces invariant under the permutation group: (1) a
(three-dimensional) symmetric subspace (spanned by three vectors:
$\{ \phi \otimes \phi , \psi \otimes \phi , \phi \otimes \phi +
\psi \otimes \phi\}$) corresponding to bosons; and (2) a
(one-dimensional) anti-symmetric subspace (spanned by one vector:
$\{ \phi \otimes \phi - \psi \otimes \phi\}$) corresponding to
fermions.\footnote{These `spanning' vectors
    correspond, of course, to the possible outcomes in the quantum
    statistics.} The symmetric subspace is quite clearly reducible, but the
    three subspaces spanning it are one-dimensional and, therefore,
    irreducible: they contain no permutation invariant proper subspaces.
    Hence, the irreducible representations correspond to
    \emph{types} of particle.\footnote{The symmetry properties
    mentioned earlier mean also that symmetry type is
    conserved; that is, state
    vectors remain in one or the other
    subspace over time: once a boson (fermion) always a boson
    (fermion)! However, this rule breaks down in supersymmetric theories,
    since such theories possess a symmetry relating Fermi (matter) to
    Bose (force); however, we shall ignore this complication here.}

    However, when we consider more than two particles (giving a non-abelian
    permutation group) we
    find that we get more than the two symmetry types that we
    observe in nature. For instance, for three
    indistinguishable particles we have 3! irreducible
    representations of the permutation group in the joint Hilbert
    space. In
    addition to the standard symmetry types (bosons and fermions), we
    also obtain `parabosons' and `parafermions', transforming differently
    under the action of the permutation group, and leading to
    alternative kinds of
    statistical behaviour known as `parastatistics'.\footnote{The
    statistical behaviour
    of these `higher-dimensional' irreducible representations is best modelled by
    the `braid group' rather than the `permutation group'.}

    Thus, with the framework we have built up so far we can see that
    there is more `mathematical structure' than there is `physical
    structure': nature has shown us (so far) that there are bosons and
    fermions,
    whilst the theory allows for particles with different
    symmetry types (with potentially observable differences).
    In order to overcome this problem Messiah (1962, 595)
    introduced a postulate, the ``symmetrisation postulate'' [SP],
    which served to restrict the possible particles to those two
    classes that we have so far found the world to be grouped into.
    This postulate can be stated simply: States of identical particle
    systems must be either symmetrical or anti-symmetrical under the action
    of permutation operators. We can now turn to the philosophical
    implications of these ideas.


\begin{center}
\textbf{3. The Relationship Between Permutation Invariance and the
Symmetrisation Postulate}
\end{center}

Let us consider in a little more detail the relationship between
SP and PI. The former is obviously a restriction on the states of
the assembly. If the latter is likewise understood, then it is
easy to see that SP is sufficient but not necessary for PI.
Understood in this way, the fact that PI is implied by SP means it
picks up indirect empirical support from the latter (assuming that
all known particles are either bosons or fermions - an assumption
which has been questioned (see the papers in Hilborn and Tino
2000)).

    However, as Greenberg and Messiah argued, PI should be
    interpreted, not as a restriction on the states, but rather
    as a restriction on the possible observables for the assembly
    (Greenberg and Messiah 1964). On this view, PI dictates
    that any permitted observable must commute with any permutation
    operator and this in turn implies that the observable must be a
    symmetric function of the particle labels. The difference
    between SP and PI can thus be expressed as follows: SP expresses
    a restriction on the states for all observables, Q; whereas PI
    expresses a restriction on the observables, Q, for all states.
    From this perspective what PI does is restrict the accessibility
    of certain states, such that once in a certain set of states,
    whether Bose-Einstein, Fermi-Dirac or parastatistical of a given
    order, the particles cannot move into a different set. However,
    the question as to its status now becomes acute. Before we
    discuss this question in more detail, we shall consider one
    further aspect of the formal representation of quantum
    statistics and PI.


\begin{center}
\textbf{4. Permutation Invariance and the Topological Approach to
Particle Identity.}
\end{center}

As is well known, Schr\"{o}dinger's early attempt to give a
broadly classical interpretation of the new quantum mechanics
foundered on the point that the appropriate space for a
many-particle wave function had to be multi-dimensional. Even
then, as Einstein pointed out, use of the full configuration space
formed by the N-fold Cartesian product of three-dimensional
Euclidean space appeared to conflict with the new quantum
statistics insofar as within this full space, configurations
related by a particle permutation are regarded as distinct. The
standard resolution of this problem, of course, is to move to the
reduced quotient space formed by the action of the permutation
group on the full configuration space, in which points
corresponding to a permutation of the particles are identified,
and then apply appropriate quantum conditions (see Leinaas and
Myrheim 1977). In this context PI is effectively coded into the
topology of configuration space itself, and the different
statistical types then correspond to different choices of boundary
conditions on the wave function ( see, for example, Bourdeau and
Sorkin 1992).

    This reduced configuration space is not in general a smooth
    manifold since it possesses singular points where two or more
    particles coincide. This leads to two technical difficulties:
    first, it is not clear how one might define the relevant
    Hamiltonian at such singularities (ibid., 687) and secondly,
    the existence of these singular points is not compatible with
    Fermi-Dirac statistics. The obvious, and now standard,
    solution is to simply remove from the configuration space the
    subcomplex consisting of all such coincidence points,
    yielding a smooth manifold. The relevant group for n particles
    is then the n-string braid group as we noted above, and the
    irreducible unitary representations of this group can be used
    to label the different statistics that are possible. Imbo,
    Shah Imbo and Sudarshan have provided a definition of the
    `statistical equivalence' of two such representations in
    terms of which they obtain not merely ordinary statistics,
    parastatistics and fractional or anyon statistics but more
    exotic forms which they call `ambistatistics' and
    `fractional ambistatistics' (Imbo, Shah Imbo and Sudarshan
    1990).
    The deployment of the braid group in this manner may appear to
    conflict with
the suggestion that one of the advantages of the configuration
space approach is that it actually excludes the possibility of
non-standard statistics. This claim is based on the work of
Leinaas and Myrheim (op. cit.) which apparently demonstrates that
for a space of dimension 3 or greater only the standard statistics
are possible. The conclusion drawn is that ``... the (anti-)
symmetrisation condition on the wave function is now seen to be
related to the dimensionality of space, in contrast to the Messiah
and Greenberg analysis wherein the (anti-) symmetrisation
condition receives the status of a postulate'' (Brown et al. 1992,
230).

 It turns out,
    however, that this treatment assumes the standard
    quantisation procedure which incorporates one-dimensional
    Hilbert spaces only. From the group-theoretical perspective
    this amounts to allowing only one-dimensional representations
    and so it should come as no surprise that para- and
    ambi-statistics cannot arise. Effectively what
    Leinaas and Myrhiem have done is to ignore the `kinematical
    ambiguity' inherent in the quantisation procedure which
    derives from the (mathematical) fact that the set of irreducible
    representations of the permutation group contains not just the
    trivial representation manifested above but also others
    corresponding to exotic statistics (Imbo et al. 1990,
     103-104).
    In what follows we shall occasionally return to the topological
    approach to see if it can shed any light on the issue of the
    status of PI.


\begin{center}
\textbf{5. Permutation Invariance and the Metaphysics of
Individuality.}
\end{center}


\begin{center}
\textbf{5.1 The Received View: Quantum Particles as
Non-Individuals.}
\end{center}

One well known approach to this issue takes PI to be profoundly
related to the peculiar metaphysical character of quantum
particles, namely that they are `identical' or `non-individual',
in some sense. Referring back to our illustration of the
difference between classical and quantum statistics above, the
argument for such a view goes like this: in classical
Maxwell-Boltzmann statistics, a permutation of the particles is
taken to give rise to a new, countable arrangement. Since the
particles are indistinguishable, in the sense of possessing all
intrinsic or state-independent properties in common (that is,
properties such as rest mass, charge, spin, etc.), this generation
of new arrangements must reflect something about the particles
which goes beyond their intrinsic properties, something which
allows us to treat them as distinct individuals. In the quantum
case - whether Fermi-Dirac or Bose-Einstein - the distribution is
permutation invariant and a permutation of the particles does not
yield a new arrangement. Hence the statistical weight in quantum
statistics - of either form - is appropriately reduced. Since the
particles are regarded as indistinguishable in the same sense as
their classical counterparts, this reduction in the count, due to
PI, must reflect the fact that the particles can no longer be
regarded as individuals - they are, in some sense,
`non-individuals'. In other words, according to this argument, PI
implies non-individuality.

    We shall call this view - that quantum particles are, in some
    sense, not individuals the Received View. It became fixed in
    place almost immediately after the development of quantum
    statistics itself (and in its modern incarnation it can be
    found in Dieks 1990, for example). Thus at the famous
    Solvay Conference of 1927, Langevin noted that quantum particles
    could apparently no longer be identified as individuals and
    that same year, both Born and Heisenberg insisted that quantum
    statistics implied that the ``individuality of the corpuscle is
    lost'' (Born 1926; see Miller 1987, 310). Some
    years later, in 1936, Pauli wrote to Heisenberg that he
    considered this loss of individuality to be ``... something
    much more fundamental than the space-time concept'' (see von
    Meyenn 1987, 339).


\begin{center}
\textbf{5.2 Challenges to the Received View.}
\end{center}

The Received View has been challenged on a variety of grounds over
the past fifteen years or so. These challenges come at the issue
from two directions, and in both cases it is denied that there
exists some fundamental metaphysical difference between quantum
and classical particles. The first challenge insists that
classical particles, like their quantum counterparts, are not only
`indistinguishable', in the above sense, but should also be
understood as subject to PI. The grounds for this claim rest on a
positivistic understanding of the meaning of `non-individuality'
which takes the latter notion to be determined experimentally
(Hestenes 1970). The idea is that non-individuality follows from
the requirement that in order for the entropy be extensive, the
relevant expression must be divided by N!, where N is the number
of particles in the assembly. It is this extensivity of entropy
which, it is claimed,  resolves the infamous Gibbs' paradox: if
like gases at the same pressure and temperature are mixed, then
there is no change in the experimental entropy. This is in
disagreement with the result obtained from Maxwell-Boltzmann
statistics, incorporating the considerations of particle
permutations sketched above. Excluding such permutations from the
calculation of the statistical entropy by dividing by N! is then
understood as resolving the `paradox'. If this were correct, then
it is claimed - classical statistical mechanics would have to be
regarded as permutation invariant also (see, for example, Saunders
forthcoming, 22, fn. 13) and the contrast with quantum physics
would have to be sought elsewhere.

    Historically, however, the failure of extensivity and the
    Gibbs Paradox were seen as revealing a fundamental flaw in
    the Maxwell-Boltzmann definition of entropy and one which
    is corrected by shifting to an understanding of the particles
    as quantum in nature.\footnote{Here we are following Post
    who writes, ``... the flaw in classical statistical mechanics
    represented by Gibbs' paradox points to a radical theory of
    non-individuality such as Bose's'' (1971, 23, fn. 50).
    By `flaw' here Post means a `neuralgic point' which can act
    as the heuristic stimulus for new theoretical development.}
    In other words, the force of the argument can be turned
    around: what it shows is that the world is actually quantum
    in nature, as one would expect. What the exclusion of the
    permutations (by dividing by N!) is a manifestation of is
    precisely that the particles are not just indistinguishable
    in the classical sense. If this aspect is incorporated into
    the analysis from the word go, the so-called `paradox' simply
    does not arise\footnote{Historically, the N! division was the
    subject of a vigorous dispute between Planck, who defended
    the move, and Ehrenfest, who argued that it was ad hoc as
    it stood and hence required further justification.}.

    The second challenge suggests that, contrary to the Received
    View quantum particles, just like their classical counterparts,
    can also be regarded as individuals. The question immediately
    arises: if this were the case, how would one account for the
    different treatment of permutations; that is, how would one
    account for the difference between classical and quantum
    statistics? We recall that this difference lies in the drop
    in statistical weight assigned to the relevant arrangements.
    This can be accounted for, without appealing to the supposed
    `non-individuality' of the particles, by focussing explicitly on
    the role of PI, understood as a kind of initial
    `accessibility' condition (French 1989a).

    To see this, let us recall that, understood as a restriction on
    the observables, PI acts as a super-selection rule which divides
    up the relevant Hilbert space into a number of irreducible
    sub-spaces, corresponding to irreducible representations of
    the symmetric group. It can be shown that transitions between
    such subspaces are (generally) forbidden (at least in
    non-supersymmetric theories). Hence PI imposes a restriction
    on the states of the assembly such that once a particle
    is in a given subspace, the other - corresponding to other
    symmetry types - are inaccessible to it. Returning to the
    argument above, the reduction in statistical weight is now
    explained by the inaccessibility of certain states, rather
    than by a change in the metaphysical nature of the particles.
    In the simple case of two particles distributed over two
    one-particle states, the only subspaces available are the
    symmetric and antisymmetric and so only one of the two
    possible states formed by a permutation is ever available to
    the system. Thus the statistical weight corresponding to the
    distribution of one particle in each such state is half the
    classical value.

    We shall examine the two components of this alternative to
    the Received View in a little more detail. First of all,
    with regard to the role of PI, the idea of restrictions on
    the set of states accessible to a system can also be found
    in classical statistical mechanics, of course. There it is
    the energy integral which imposes the most important
    restriction as it determines which regions of the relevant
    phase space are accessible. Other uniform integrals of the
    motion may also exist for a particular assembly but these
    are not generally thermodynamically significant. What PI
    represents is an additional constraint or initial condition,
    imposed on the situation. In particular, the symmetry type of
    any suitably specified set of states is an absolute constant
    of motion equivalent to an exact uniform integral in classical
    terms (see Dirac op. cit., 213-216). Of course, some may
    wonder whether this actually sheds much light on the status
    of PI, since it appears to leave it standing as a kind of
    `brute fact' but at least it is no more brutish than the other,
    classical, constraints.

    Secondly, there is the issue of how we are to understand the
    individuality of the particles.  As is well known, we have a
    range of options to choose from, some more attractive than
    the others:

\vspace{5pt}

(1) Haecceity or primitive thisness (Adams 1979).

(2) Some form of Lockean substance.

(3) Spatiotemporal location.

(4) Some subset of properties.

\vspace{5pt}

\noindent The first two are perhaps the least attractive as far as
broadly `empiricist' philosophers are concerned, since they appeal
to factors which are utterly empirically superfluous (see French
and Redhead 1988; Redhead and Teller 1991 and 1992; van Fraassen
1991). However, they share what some would see as the advantage of
making manifest the conceptual distinction between individuality
and distinguishability, where the latter is to be understood in
terms of some difference in properties.  The third and fourth
collapse this distinction by taking that which renders the entity
distinguishable as that which also `confers' individuality. Each
requires some extra postulate, however, in order to effect this
collapse. In the case of option (3), this extra `something' is the
postulate that the particles are impenetrable, since if they were
not, their spatiotemporal locations could not serve to distinguish
and hence individuate them. As is well known, however, this option
is problematic in the quantum context, as standardly understood,
since it can be shown that the family of observables corresponding
to the positions of single particles cannot provide distinguishing
spatiotemporal trajectories (as Huggett and Imbo 2000 emphasise,
one can prove the more general result that no family of
observables can provide such trajectories)\footnote{They also
stress that this lack of trajectories does not imply anything
about the status of PI.}.

    Nevertheless, considerations of impenetrability do feature
    within the topological approach.  We recall that, according
    to this perspective, we must move to the reduced configuration
    space formed by removing the points where the particles would
    coincide. An obvious justification for this adverts to the
    impenetrability of the particles, understood, in turn, as due
    to certain repulsive forces holding between them. Such a
    conjecture has been made in the case of anyons (see Aitchison
    and Mavromatos 1991) and more generally this has been
    taken to confer a further advantage on the configuration space
    approach, in the sense that

\begin{quotation}
... it allows particle statistics to be understood as a kind  of
`force' in essence similar to other interactions with a
topological character, like the interaction between an electric
and magnetic charge in three spatial dimensions, or the type of
interaction in two dimensions which is responsible for the
Bohm-Aharonov effect and fractional statistics. (Bourdeau and
Sorkin op. cit., 687).\footnote{Again, we can't help but recall
some relevant history here. The suggestion that the non-classical
aspects of quantum statistics reflects a lack of statistical
independence and hence a kind of correlation between the particles
can be traced as far back as Ehrenfest's early reflections on
Planck's work and crops up again and again in the literature. On
the philosophical side, Reichenbach (1956, 234-235) argued that
such correlations - taken realistically in this sense - represent
causal anomalies in the behaviour of the particles: for bosons
these anomalies consist in a mutual dependence in the motions of
the particles which could be characterised as a form of
action-at-a-distance; for fermions, the anomaly is expressed in
the Exclusion Principle if this is interpreted in terms of an
interparticle force. As far as Reichenbach himself was concerned,
such acausal interactions should be rejected and thus he preferred
the account of quantum statistics which emphasises the
metaphysical lack of individuality of the particles and in which
the correlations are not regarded in force-like terms.}
\end{quotation}

    Some have regarded such an understanding as suspiciously ad
    hoc (see Brown et al., op. cit.). One way of eliminating
    this `ad hocness' is to shift to the framework of de Broglie-Bohm
    pilot wave theory\footnote{Of course, adopting such a framework
    means abandoning the standard eigenvalue-eigenstate link and the
    latter is precisely what is assumed in the above proofs that
    spatiotemporal trajectories cannot serve to distinguish.}. Here, as is
    well known, there is a dual
    ontology of point particles plus pilot wave, where the role of
    the latter is to determine the instantaneous velocities of the
    former through the so-called `guidance equations' (ibid.). Since
    these equations are first-order, the trajectories of two particles
    which are non-coincident to begin with will never coincide.
    In effect the impenetrability of the particles is built into
    the guidance equations and the singularity points remain
    inaccessible. Hence the conclusion that ``... within the
    topological approach to identical particles the removal of
    the set ... of coincidence points from the reduced configuration
    space ...  follows naturally from de Broglie-Bohm dynamics as
    it is defined in the full space ...'' (ibid., 233)\footnote{There
    is the worry that this might exclude the possibility of
    Bose-Einstein statistics, if it is the case that the de
    Broglie-Bohm trajectories never in fact cross. Brown et. al.
    argue that this latter claim is simply not correct since
    symmetry considerations demonstrate that if the particles
    coincide at all, they coincide forever. If the bosons are
    initially separated, then the relative velocity vanishes at
    the coincident point which together with the first order
    nature of the guidance equations means that they can never
    cross (op. cit., 233).}.

    It is part of the attraction of this framework that it retains,
    or appears to retain, a form of classical ontology which meshes
    well with the metaphysical view of particles as
    individuals\footnote{Brown et. al. explicitly note this point.
    Nevertheless it is not clear that the particles can in fact be
    regarded straightforwardly as individuals in the classical sense
    (see French 2000). Further criticisms of the view that
    de Broglie-Bohm theory is philosophically classical can be
    found in Bedard 1999. For a response see Dickson
    2000.}. However, it is also important to recognise
    that the topological approach can also accommodate the Received
    View of particles as non-individuals. Indeed, one of the
    motivations given by Leinaas and Myrheim is that it allows
    one to dispense with the whole business of introducing
    particle labels and then effectively emasculating their
    ontological force by imposing appropriate symmetry constraints
    (op. cit., 2; cf. also Bourdeau and Sorkin op. cit., 687).
    Of course, if one is going to insist that the particles are
    non-individuals, then some alternative justification for the
    removal of the coincidence points must be sought for. One
    possibility is to tackle the problem of collisions directly.
    Bourdeau and Sorkin, for example (op. cit.), show that for
    fermions, the self-adjoint extension of the Hamiltonian to
    cover the singularities is unique, at least in the
    two-dimensional case, so that collisions are strictly forbidden,
    whereas in the case of both Bose-Einstein and fractional
    statistics there are a range of alternative extensions, some
    of which allow collisions but some which do not\footnote{Thus
    they argue that simply cutting out the singular points results
    in a loss of information.}. By requiring that the wave-function
    remains finite at the coincident point they argue that a
    unique choice of Hamiltonian can then be made and it turns
    out that collisions are allowed only in the case of Bose-Einstein
    statistics. Thus, whereas for fermions it doesn't really matter
    whether the singular points are retained or not, for bosons and
    anyons, on this account, it does, since these points are either
    the locations of collisions in the boson case or the locations
    of vanishing $\Psi$ for anyons.


\begin{center}
\textbf{6. Individuality and the Identity of Indiscernibles.}
\end{center}

Let us return to our list of options for understanding quantum
individuality. Option (4) attempts to ground it in some subset of
properties of the particles. This also requires a supplementary
principle in order to block the possibility of two individuals
sharing the same subset of properties, and this is provided, of
course, by the Principle of Identity of Indiscernibles (PII). In
terms of second-order logic with equality, PII can be written as

\begin{equation}
\forall \Gamma \{ \Gamma (a) \equiv \Gamma (b) \} \rightarrow a =
b
\end{equation}

\noindent where `a' and `b' are individual constants designating
the entities concerned and $\Gamma$ is a variable ranging over the
possible attributes of these entities. Different forms of PII then
arise depending on what sort of attributes feature in the range of
$\Gamma$\footnote{We exclude the attribute of `being identical
with a', since PII would then simply be a theorem of second-order
logic. Furthermore, Adams identifies haecceity or `primitive
thisness' with precisely this attribute (op. cit.) and hence
admitting it here would be tantamount to adopting option (1).}.
The logically weakest form, PII(1), states that it is not possible
for two individuals to possess all properties and relations in
common; PII(2) excludes properties and relations which can be
described as spatiotemporal; while the strongest form PII(3),
includes only monadic, non-relational properties. Before we
consider the status of these forms of PII in quantum physics, it
is worth noting, first of all, that PII(1) has often been taken as
necessarily true on the grounds that no two individuals can
possess exactly the same spatiotemporal properties or enter into
exactly the same spatiotemporal relations (see, for example,
Quinton 1973, 25). This obviously assumes that the individuals
concerned are impenetrable and amounts to a form of option (3)
above. Both PII(1) and PII(2) allow for the possibility that
relations might be capable of distinguishing entities and hence
confer individuality (see for example Casullo 1984, who argues
that the view of entities as nothing more than bundles of
properties and relations is only plausible if based on PII(1) with
relations given the capacity to individuate). However, such a
possibility has been vigorously disputed on the grounds that since
relations presuppose numerical diversity, they cannot account for
it (see Russell 1956 and Armstrong 1978, 94-95). We shall return
to this possibility shortly.

    When it comes to the status of PII in quantum physics, if the
    non-intrinsic, state-dependent properties are identified with
    all the monadic or relational properties which can be expressed
    in terms of physical magnitudes associated with self-adjoint
    operators that can be defined for the particles, then it can
    be shown that two bosons or two fermions in a joint symmetric
    or anti-symmetric state respectively have the same monadic
    properties and the same relational properties one to another
    (French and Redhead 1988; see also Butterfield
1993). On the basis of such an identification,  even the weakest
form of the Principle, PII(1), fails for both bosons and fermions
(French and Redhead op. cit.; French 1989b).\footnote{Margenau had
earlier concluded that PII(3) fails, since the same reduced state
can be assigned to the fermions in an antisymmetric state and
hence they possess the same monadic properties (Margenau 1944).
This conclusion has been criticised on the grounds that such
reduced states cannot be regarded either as ontologically separate
or as encoding genuinely monadic properties (see Mittelstaedt and
Castellani 2000; Massimi 2001).} Hence the Principle of Identity
of Indiscernibles cannot be used to effectively guarantee
individuation via the state-dependent properties and option (4)
fails.\footnote{For alternative discussions see Cortes 1976;
Barnette 1978; Ginsberg 1981; Teller 1983; and van Fraassen 1985
and 1991.}, leaving Lockean substance or primitive thisness as the
only alternatives

    However, there may still be hope for this option. Saunders
    (ibid., 10-11)  has recently revived Quine's proposal for the
    analysis of identity, which he understands as yielding a version
    of PII. Roughly speaking this is the condition
    that $x = y$ (where $x, y, u_{1}, u_{2}$, ... are variables)
    if and only if, for all unary predicates $A$, binary
    predicates $B$, ..., $n$-ary predicates $P$, we have

\begin{itemize}
\item $A(x) \equiv A(y)$

\item $B(x, u_{1}) \equiv B(y, u_{1}); B(u_{1}, x) \equiv B(u_{1},
y)$

\item $P(x, u_{1}, ..., un_{i1}) \equiv B(y, u_{1}, ..., un_{i1})$
and permutations
\end{itemize}

\noindent together with all universal quantifications over the
free variables $u_{1}, ..., un_{i1}$ other than x and y (ibid.,
11). If the relevant language contains monadic predicates only,
then this principle amounts to the claim that two entities are
identical if and only if they have all properties in common. Two
entities are said to be absolutely discernible if there is a
formula with only one free variable which applies to one entity
but not the other. With only monadic predicates allowed, the
principle states that numerically distinct entities are absolutely
discernible. If relations are admitted, however, one can have
entities which are not identified by the principle yet are not
absolutely discernible. Two entities are said to be relatively
discernible if there is a formula in two free variables which
applies to them in any order. But there is a further category: If
the admitted relations include some which are irreflexive, then
one can have entities which are counted as distinct according to
the principle but are not even relatively discernible. These are
said to be weakly discernible, and the principle excludes the
possibility of entities neither absolutely, relatively, or weakly
discernible.

    This, it is claimed, is more natural from a logical point of view
    (being immune to the standard counter examples - such as the
    infamous two globes - which beset PII; see Saunders (ibid., 7),
    and is also better suited to quantum mechanics in that, unlike
    traditional versions of  PII,
     it is not violated by fermions at least, since an irreflexive
     relation always exists between them. Consider for example, two
fermions in a spherically-symmetric singlet state. The fermions
are not only indistinguishable but also have exactly the same
spatiotemporal properties and relations in themselves and
everything else. However, each satisfies  the symmetric but
irreflexive relation of "having opposite direction  of spin to"
and so are weakly discernible. Thus for fermions, at least, we
have the possibility of grounding their individuality via a
version of PII, without having to appeal to anything like
primitive thisness.\footnote{No such possibility exists for
bosons; however, Saunders adopts the Redhead and Teller option of
regarding them as non-individual field quanta.  He takes this
metaphysical difference as tracking the physical one between the
`stable constituents of ordinary matter' (fermions) and gauge
quanta (bosons), although it is not clear why the metaphysics
should follow the physics in this particular way, or at all.}
    However, there is an obvious concern one might have here, which
    reprises the worry hinted at above, regarding the individuating
    power of relations: doesn't the appeal to irreflexive relations
    in order to ground the individuality of the objects which bear
    such relations involve a circularity? Such concerns are rooted
    in the - apparently plausible - view that relata have ontologically
     priority over relations, such that the former can be said to
     `bear' the latter. Suppose we were to drop such a view. Of course,
      in order to describe the relation - either informally as above
      or set-theoretically in terms of an ordered tuple
      $\langle x, y \rangle$ - we have to introduce some form of
      label, as in the example above, but description should not
      dictate conceptualisation. The label can be understood as a
      kind of place-holder and instead of talking of relata `bearing'
      relations, one can talk of the intersection of relations as
      constituting relata, as Cassirer did, or of relata as unifying
      relations, as Eddington did. The names here give the game away -
      what this amounts to is some kind of structuralist ontology
      which allows for individuation via relations.


\begin{center}
\textbf{7. PI is Neither Sufficient nor Necessary for
Non-Individuality.}
\end{center}

    Returning now to the issue of the status of PI, the point we want
    to emphasise is that something further needs to be added to get
    from it to the Received View of particles as non-individuals. In
    other words, PI is not sufficient for non-individuality. The
    question remains,  is it necessary?

    van Fraassen, for example, has answered that it is not
    (1991, 375), while acknowledging that the claim - that
    ``when identity is properly understood, it entails Permutation
    Invariance tautologically'' (ibid.) - nevertheless contains a
    `core' of truth. Butterfield has disagreed, however, insisting
    that this claim merely summarises the motivation for PI,
    namely that expectation values for the composite system
    cannot be sensitive to the differences between $\phi$
    and $P\phi$ (1993, 457). At issue here, of course,
    is what is meant by `identity', or non-individuality,
    being `properly understood'. Butterfield's understanding
     appears to correspond to Dirac's above, but if one were
     to reject this as broadly positivistic, how do the
     alternatives stand up?

    In order to examine this question, we need to consider
    what we mean by non-individuality in this context. All of
    the options considered here are constructed through a
    combination of indistinguishability - of course - together
    with the denial of some `principle' of individuality, whether
    that be substance, primitive thisness, spatiotemporal
    trajectories, or PII. The question then is whether any of
    these alternatives can provide the restriction on observables
    that PI demands. Clearly neither substance nor primitive
    thisness can, since by their very nature, neither can be
    expressed in terms of observables! What about indistinguishability
    plus the denial of spatiotemporal trajectories? Huggett
    and Imbo (2000) have recently considered this case and
    have argued that here too PI does not follow. Their argument
    considers each conjunct separately: first of all, the absence
    of spatiotemporal trajectories follows from the dynamics of
    QM  together with the general way observables are treated and
    again this imposes no restrictions on these observables. Hence,
    the lack of spatiotemporal trajectories is simply irrelevant
    in this case. Furthermore, there exists the possibility of
    particles, such as first quantised versions of Greenberg's
    quons (Greenberg 1991)\footnote{This possibility is
    not unproblematic, of course. The q-mutator formalism
    apparently requires some observables to be non-local and
    there is the further issue as to whether this possibility
    is ruled out on experimental grounds (again see the papers
    in Hilborn and Tino 2000).} which are indistinguishable
    but for which every Hermitian operator is an observable -
    hence some of the observables are non-symmetric and PI is violated.

    Of course, if PII were to hold for quons, then the implication
    would be restored for that particular understanding of
    non-individuality; that is, if quons could not be considered
    as non-individuals in this sense, then the violation of PI
    would not be indicative of the failure of the entailment.
    However, what would have to be shown is that there are no
    possible quon states for which PII is violated in the same
    ways as for bosons and fermions. In the case of paraparticles,
    French and Redhead 1988 showed that although there do
    exist states for which the monadic properties of all the
    separate particles are the same, there also exist possible
    paraparticle states for which PII is violated. So far as we
    know, nothing equivalent has been demonstrated for quons,
    although in what little philosophical discussion there has
    been about them, it appears to be assumed that such PII
    violating states exist (Hilborn and Yuca forthcoming).

    The choice, then, is stark: either adopt the Dirac/Butterfield
    understanding of non-individuality, in which case PI is indeed
    an expression of it but not in a metaphysically interesting way,
    or accept that non-individuality does not imply permutation
    invariance. The latter option leaves PI metaphysically ungrounded.


\begin{center}
\textbf{8. Underdetermination and the Structuralist View of
Particles}
\end{center}

This metaphysical `detachment' of PI has been expressed in terms
of a kind of underdetermination which holds between the Received
View, in which PI is tied up with the non-individuality of the
particles and the alternative account of particles as individuals,
with PI taken as some sort of initial condition (French and
Redhead op. cit.). As we have indicated, PI does not discriminate
between these conceptual possibilities (French 1989a; 1998; van
Fraassen 1991; Huggett 1995; Balousek 2000) and hence any argument
for one or the other is going to have to proceed on different
grounds. Most famously, perhaps, Redhead and Teller have
elaborated a methodological argument to the effect that the
Received View meshes better with the metaphysics of Quantum Field
Theory where - it is claimed - individuality is not assumed from
the word go (this argument goes back to Post 1963). This last
claim has been disputed (by de Muynck 1975 and van Fraasssen 1991;
for criticisms of the latter, see Butterfield 1993) and Balousek
has insisted that to appeal to methodology to break the
underdetermination is to accede to a form of conventionalism
(Balousek 2000; for a response see Teller and Redhead 2001).

We shall not pursue the ins and outs of this debate here. The
alternative is to accept the underdetermination and explore its
implications. As far as the particles themselves are concerned, it
motivates a shift - some might say retreat - to a structuralist
view of entities which eschews talk of individuality or
non-individuality entirely (see Ladyman 1998). According to such a
view, the particles are nothing but `nodes' or `intersections' in
some kind of physical structure, which now bears all the
ontological weight. We shall consider how PI looks from such a
position shortly.


\begin{center}
\textbf{9. The Experimental and Theoretical Status of PI.}
\end{center}

Returning to the issue of the status of PI, let us  consider
whether there could there be direct evidence for the principle. We
can examine this question in a broader context: what is required
for any symmetry principle to be observable? Kosso has argued that
the distinction between observable and unobservable symmetries
matches that between global and local. Thus Lorentz invariance,
for example, is directly observable whereas general covariance is
not, since a specific dynamical principle  such as the principle
of equivalence  must be assumed in order to infer the symmetry
from the observation (Kosso 2000). Now, a problem arises when it
comes to symmetries such as PI: in order to observe whether a
symmetry holds or not, one must be able to observe that a) the
specified transformation has taken place; and b) that the
specified invariant property remains the same under the
transformation (ibid., 86). The first condition requires there to
be a fixed point of reference with respect to which the
transformation can be measured. In the case of the permutation of
protons and neutrons underlying isospin symmetry, for example,
there exists a `fixed standard' of what it is to be a neutron and
what it is to be a proton with respect to which the permutation
makes sense (ibid.). If, however, the symmetry is `observationally
complete', in the sense that all of the observable properties of
the system are invariant under the transformation, then the
symmetry will be unobservable in principle (ibid., 88). Kosso does
not consider PI in his discussion but it is observationally
complete in precisely this sense: in order to test whether it
holds or not, one would have to be able to experimentally
distinguish the states represented by$\mid \Psi \rangle$ and
$\hat{P}\mid \Psi \rangle$ to begin with. However, that is
precisely what the principle itself denies (Balousek 1999, 20).

    If there can be no direct evidence for PI, is it demanded by
    the theory of quantum mechanics itself?  It is often claimed
    that PI is not logically required by the axioms of the theory
    (Balousek ibid., 20), but of course, this depends on what
    the latter are taken to include. Adopting a crude historical
    perspective, the work of Weyl, Dirac and von Neumann can be
    seen as an attempt to impose order upon what was, in the
    late 1920s, a bit of a hodgepodge of laws and principles,
    some of them `phenomenological' in nature, such as the
    Exclusion Principle. Weyl's framework was explicitly
    group-theoretical and here, as we have indicated, PI can be
    incorporated within this framework as an expression of the
    metaphysical nature of quantum objects. Dirac, on the other
    hand, eschewed both group theory and metaphysics, preferring
    his own `bra' and `ket'  framework in which PI is seen as
    nothing more than an expression of the observational
    indistinguishability of $\mid \Psi \rangle$ and $\hat{P}\mid
    \Psi \rangle$, from a rather crude verificationist perspective.
    In the context of von Neumann's Hilbert space formalism PI does
    appear to be an extra postulate reflecting either the
    metaphysical nature of quantum particles or some kind of
    `initial condition' as we noted above but of course one
    could make the case that it should be added to the standard
    axioms of QM whatever they are  in order to extend the
    theory to give a quantum statistics. Whatever framework one
    chooses, claims that PI is, in some sense, `ad hoc' must be
    treated with caution.
    Where does all this leave the status of PI? It appears to be
    a kind of `free-floating' principle, one that is required
    neither experimentally nor metaphysically.
    Huggett has proposed that it be regarded straightforwardly
    as a symmetry on a par with rotational symmetry, for example
    (Huggett 1999). Now, of course, as Huggett acknowledges,
    the two symmetries are very different\footnote{A quantum
    system of the kind we have been considering
    is not just \emph{covariant} with respect to permutations but
    \emph{invariant}:
    permutations are not just indistinguishable to appropriately
    transformed observers but to \emph{all} observers.}, but, nevertheless,
    PI  is implied by the conjunction of a further symmetry
    principle which space-time symmetries also obey together
    with the formal structure of the permutation group. This
    further principle is what he calls `global Hamiltonian symmetry'
    which implies that the relevant symmetry operator commutes
    with the relevant Hamiltonian\footnote{What we take the
    relevant Hamiltonian to cover is crucial here because, again
    as Huggett acknowledges, the principle would appear to be
    violated in the case where, for example, we have a noncentral
    potential term in the Hamiltonian of an atomic system, but, he
    insists, the symmetry is restored if we consider the `full'
    Hamiltonian of system plus field, which does commute with the
    operators of the rotation group. As he points out (ibid., 345),
    if observers are taken to be systems too, this symmetry
    principle is equivalent to covariance for space-time
    symmetries.}. With regard to the permutation group,
    of course, permutations of a sub-system are permutations
    of the whole system and this `global Hamiltonian symmetry'
    very straightforwardly implies PI, without any additional
    assumptions concerning the structure of state space
    (ibid., 344-345).\footnote{It does, however, assume that the
    system being measured and the measurement apparatus are composed of
    the same indistinguishable particles, otherwise the Hamiltonian
    will not remain unchanged. Thanks to Nick Huggett for pointing
    this out.} Hence, Huggett concludes,

\begin{quotation}
... we should view permutations in a similar light to rotations:
we should not take [permutation invariance] as a fundamental
symmetry principle in order to explain quantum statistics. Instead
we should recognize that it is a particular consequence of global
Hamiltonian symmetry given the group structure of the
permutations. Further, if we accept the similarity of permutation
and rotation symmetry, it becomes natural to see quantum
statistics as a natural result of the role symmetries play in
nature. (ibid., 346).
\end{quotation}

    However, as Huggett acknowledges, permutation invariance only
    follows from his general symmetry principle given the particular
    structure of the permutation group. So the issue of the status
    of PI is pushed back a step: what is the status of the structure
    of the permutation group? Or, to put it another way, why should
    that particular group structure be applicable?
    At one extreme we have the view that it is \emph{a priori}.
    As is well known, Weyl insisted that ``all \emph{a priori} statements
    in physics have their origin in symmetry'' (1952, 126).
    Not surprisingly, empiricists such as van Fraassen have tended
    to resist this line (van Fraassen 1989) and move to
    the other end of the spectrum, offering a broadly pragmatic
    answer to our question. From this perspective, PI comes to be
    seen as nothing more than a problem solving device
    (see Bueno 2001). Occupying the middle ground we
    have the following alternative answers to our question:

\begin{enumerate}
 \item It is just a brute fact. We have already encountered
this option in our discussion of PI as an `initial condition'
imposed on the situation.

\item It is to be understood as reflecting the metaphysically
peculiar nature of the particles themselves. However, given that
the particles can also be described  in a metaphysically
straightforward way - as individuals - this option is always going
to require some further principle whose status may be less well
grounded than that of PI itself\footnote{This is, in essence, the
heart of the dispute between Balousek and Redhead and Teller.}.

\item It is to be understood as reflecting a structural aspect of
the world. From this perspective, that the permutation group is
applicable is neither a simple `brute fact' nor metaphysically
derivative, in the sense of mathematically representing the nature
of the particles, but rather it represents something profoundly
structural about the world.

\end{enumerate}

Huggett rejects the first two options but then leaves the
metaphysical status of PI hanging. We shall pursue the last option
a little further in another context, namely the connection between
permutations of particles and diffeomorphisms of space-time
points.


\begin{center}
\textbf{10 Permutation Symmetry, Structuralism and Diffeomorphism
Invariance.}\label{S:di}
\end{center}

Structuralism has a long and interesting history which is
intimately bound up with developments in physics. Both General
Relativity and Quantum Mechanics had a profound impact on the work
of early structuralists such as Cassirer and Eddington (for
discussion, see French 2001). Putting it crudely, the central idea
of this programme is to effectively deconstruct the `object of
knowledge' - whether space-time or quantum particles - into a web
of relations bound together by symmetry principles represented
group theoretically. If we focus on the group-theoretic
representation of invariant properties such as mass and spin, what
this `deconstruction' yields are kinds of particles (see
Castellani 1993 and 1998). Similarly, PI can be seen as embodying
a form of structuralist representation of `broader' kinds, namely
bosons, fermions, parabosons, parafermions and so on. In other
words, the status of PI, from this perspective, is that of one of
the fundamental symmetry principles which effectively binds the
`web of relations' constituting the structure of the world into
these broad kinds.


\begin{center}
\textbf{10.1 Permutation Symmetry and Structural Realism.}
\end{center}

It is this kind of structuralist deconstruction of objects which
is incorporated into Ladyman's `ontic' form of structural realism,
alluded to above. As we indicated, this attempts to avoid the
metaphysical underdetermination that PI yields by
reconceptualising quantum objects entirely in structural terms
(see French and Ladyman 2002 and Saunders 2002). However, the
following objection has been raised to such a move: if structure
is understood in `relational' terms - as it typically is - then
there need to be `relata', and these cannot be relational
themselves.  The force of the objection is clearly seen in the
case of PI: we began by considering the distribution of objects
over states and the effects of permutations on such objects. How
can PI play a part in the `deconstruction' of objects into
structures when its very articulation is based on the assumption
that there are objects to begin with?
    In responding to this objection, structuralists typically appeal
    to the following manouevre (it can be found in Eddington and,
    before him, Poincar\'{e}, for example): we recall that we begin
    by introducing particle labels and it is upon these that the
    particle permutation operator acts. We then assume that these
    labels denote objects - an assumption that may be supported
    by the observation of the individual flashes on a scintillation
    screen, for example -  and this allows us to apply the
    mathematics of group theory (with its underlying standard set
    theory). However, once we have obtained the relevant structure,
    we can dispense with our original assumption, regarding it as no
    more than a heuristic crutch and the labels as simply convenient
    place holders which serve only to help us focus attention on what
    is metaphysically fundamental - PI in this case. To use a famous
    metaphor, the object is a kind of `ladder' which we use to reach
    the structure but which we can then `throw away', or `deconstruct'.
    Of course, there are other objections to structural realism which
    must be addressed (see, for example, Bueno op. cit., and
    Chakravartty 2001) but our intention here is just to
    indicate what may be a natural home for the structuralist
    understanding of PI. Furthermore, this sort of picture can
    accommodate a structuralist conception of space-time as well.


\begin{center}
\textbf{10.2 Diffeomorphisms, Permutations, and the Structuralist
Conception of Space-Time.}
\end{center}

Stachel (2002) has recently explored the connections between the
interpretation of general covariance and permutation invariance on
the one hand and the metaphysical analyses of space-time points
and quantum particles on the other. He begins by abstracting the
differentiability and continuity properties of manifolds leaving a
bare set of points. The (continuous) principle of diffeomorphism
invariance then becomes (discrete) permutation invariance. A
version of the hole argument can then be seen to apply to this set
(see Norton 1988 for an elementary account of this argument). We
have already seen how such a set, along with PI, models the
statistical behaviour of ensembles of indistinguishable quantum
particles. Thus the analogy is complete and extends, \emph{mutatis
mutandis}, to any theory which ``demands the complete
indistinguishability of its fundamental objects'' (ibid., 18).

On this basis the choice between `substantivalist' and
`relationist' conceptions of space-time is rendered as that
between the `individual' and `non-individual' metaphysics of
points. Stachel himself opts for the latter package (applying the
result to both the space-time and particle cases). The `reduced
phase space' method of solving the hole argument\footnote{This
solution takes the equivalence classes of diffeomorphic states
(i.e., the gauge orbits) as the points of a new `reduced' or
`physical' phase space.} is then understood as applying to the
permutation case (where the gauge orbits are equivalence classes
of permuted states). Since that solution is seen by Stachel as
corresponding to a relational solution, the particle case is
understood similarly. The idea is that the objects in a set (be
they the points of a manifold or the members of a quantum
ensemble) are individuated only by the relations of that set:
indistinguishable objects are not individuals intrinsically but
derive that property from relations. We have been here before, of
course, with Saunders' Quinean approach to PII above and, indeed,
Saunders also applies this approach to the space-time case.

The central idea here, then, is that any theory that demands the
complete indistinguishability of its fundamental objects \emph
{requires} invariance under the full permutation group for
discrete symmetries or the diffeomorphism group for continuous
symmetries. Stachel explicitly draws the analogy between
substantivalism in the space-time case and assuming individuality
for quantum particles, and relationism and non-individuality,
respectively. Moving in the other direction both Maidens (1993)
and Hoefer (1996) have explored the idea of regarding space-time
points as `non-individuals' in some sense (Saunders op. cit.
endorses Hoefer's conclusions). What then becomes crucial is the
\emph {sense} in which the points are regarded as non-individuals,
just as we have discussed for particles. Stachel, in particular,
understands the non-individuality of particles as their being
individuated ``entirely in terms of the relational structures in
which they are embedded'' (hence the analogy with relationism on
the space-time side). But then it is not clear what metaphysical
work the notion of `non-individuality' is doing, when we still
have `objects' which are represented by standard set theory (and
this is precisely the criticism that can be levelled against
attempts to import non-individuality into the space-time context).
What one needs to do to flesh out such an account is to apply to
space-time points the kind of `quasi-set' theory that has been
applied to non-individual quantum particles (Krause 1992). Of
course, Stachel could still maintain individuality in both cases
but at the price of introducing inaccessible states in quantum
mechanics and indeterminism in spacetime theory. In both cases we
seem to have a kind of metaphysical underdetermination.

Again the alternative, `middle way', is to drop objects out of the
ontology entirely, regarding both space-time and particles in
structural terms. Indeed, this appears to be the more appropriate
way of understanding both Stachel's talk of individuating objects
``entirely in terms of the relational structures in which they are
embedded'' (ibid., 2) and, as we have seen, Saunders' account of
``weakly discernible'' entities. However, rather than thinking of
the objects being individuated, we suggest they should be thought
of as being structurally constituted in the first place. In other
words, it is the relational structures which are regarded as
metaphysically primary and the objects as secondary or `emergent'.
The labels that appear in both cases, assigned to space-time
points and particles respectively, are just mathematical devices
which allow us to apply our set-theoretical resources.  And in
both cases, the relevant symmetries, encoded in diffeomorphism
invariance and PI, respectively, will be seen as essential
components of this structural metaphysics.

There is, for sure, plenty of work to be done here; but this
structural perspective places symmetry at the heart of our
metaphysics, and that surely makes the task a tantalising one!

\begin{center}
\textbf{References.}
\end{center}

\noindent Adams, J. (1979), `Primitive Thisness and Primitive
Identity', \emph{Journal of Philosophy} 76, 5-26.

\vspace{5pt}

\noindent Aitchison, I.J.R. and Mavromatos, (1991),
\emph{Contemporary Physics}, 32, 219.

\vspace{5pt}

\noindent Armstrong, D. (1978), \emph{Nominalism and Realism},
Vols. I and II. Cambridge University Press.

\vspace{5pt} \noindent Balousek, D. (1999, preprint),
`Indistinguishability, Individuality and the Identity of
Indiscernibles in Quantum Mechanics'.

\vspace{5pt} \noindent Balousek, D. (2000), `Statistics, Symmetry
and the Conventionality of Indistinguishability in Quantum
Mechanics', \emph{Foundations of Physics} 30,  1-34.

\vspace{5pt}

\noindent Barnette, R. L. (1978), `Does Quantum Mechanics Disprove
the Principle of the Identity of Indiscernibles?',
\emph{Philosophy of Science} 45, 466-470.

\vspace{5pt} \noindent Bedard, K. (1999), `Material Objects in
Bohm's Interpretation', \emph{Philosophy of Science} 66, 221-242.

\vspace{5pt}

\noindent Born, M. (1926), `Quantenmechanik der Stobvurgnge',
\emph{Zeitschrift f\"{u}r Physik} 38, 803-827.

\vspace{5pt} \noindent Bourdeau, M. and Sorkin, R.D. (1992), `When
Can Identical Particles Collide?', \emph{Physical Review} D 45,
687-696.

\vspace{5pt} \noindent Brown, H., Sj\"{o}qvist, E. and
Bacciagaluppi, G. (1999), `Remarks on Identical Particles in de
Broglie-Bohm Theory', \emph{Physics Letters} A 251, 229-235.

\vspace{5pt} \noindent Bueno, O. (2001), `Weyl and von Neumann:
Symmetry, Group Theory, and Quantum Mechanics',
http://philsci-archive.pitt.edu/documents/disk0/00/0\\0/04/09/index.html.

\vspace{5pt} \noindent Butterfield, J. (1993), `Interpretation and
Identity in Quantum Theory', \emph{Studies in History and
Philosophy of Science} 24, 443-476.

\vspace{5pt} \noindent Casullo, A. (1984), `The Contingent
Identity of Particulars and Universals', \emph{Mind} 123, 527-554.

\vspace{5pt} \noindent Castellani, E. (1993), `Quantum Mechanics,
Objects and Objectivity', in \emph{The Foundations of Quantum
Mechanics - Historical Analysis and Open Questions}, C. Garola and
A. Rossi. Kluwer, 105-114.

\vspace{5pt} \noindent Castellani, E. (1989), `Galilean Particles:
An Example of Constitution of Objects' in \emph {Interpreting
Bodies: Classical and Quantum Objects in Modern Physics}, E.
Castellani (ed.). Princeton: Princeton University Press, 181-194.

\vspace{5pt} \noindent Chakravartty, A. (forthcoming), `Science,
Metaphysics and Structural Realism', Cambridge University
Preprint.

\vspace{5pt} \noindent Cortes, A. (1983), `Leibniz's Principle of
the Identity of Indiscernibles: A False Principle',
\emph{Philosophy of Science} 43, 491-505.

\vspace{5pt} \noindent de Muynck, W. (1975), `Distinguishable and
Indistinguishable-Particle Descriptions of Systems of Identical
Particles', \emph{International Journal of Theoretical Physics} 14
327-346.

\vspace{5pt} \noindent Dickson, M. (2000), `Discussion: Are There
Material Objects in Bohm's Theory?', \emph{Philosophy of Science}
67, 704-710.

\vspace{5pt} \noindent Dieks, D. (1990), `Quantum Statistics,
Identical Particles and Correlations', \emph{Synthese} 82,
127-155.

\vspace{5pt} \noindent  Dirac, P. A. M. (1958), \emph{The
Principles of Quantum Mechanics}.
 Clarendon Press: Oxford.

\vspace{5pt} \noindent French, S. (1989a), `Identity and
individuality in classical and quantum physics',
\emph{Australasian Journal of Philosophy} 67, 432-446.

\vspace{5pt}

\noindent French, S. (1989b), `Why the Principle of Identity of
Indiscernibles is not Contingently True Either', \emph{Synthese}
78, 141-166.

\vspace{5pt}

\noindent French, S. (1998), `On the Withering Away of Physical
Objects', in \emph{Interpreting Bodies},  E. Castellani (ed.).
Princeton University Press, 93-113.

\vspace{5pt} \noindent French, S. (2000), `Putting a New Spin on
Particle Identity', in \emph{Spin-Statistics Connection and
Commutation Relations: Experimental Tests \& Theoretical
Implications}, R. Hilborn and G.M. Tino (eds.). American Institute
of Physics, 305-317.

\vspace{5pt} \noindent French, S. (2001), `Symmetry, Structure and
the Constitution of Objects',
http://philsci-archive.pitt.edu/documents/disk0/00/00/03/27/index.html.

\vspace{5pt}

\noindent French, S. and Ladyman, J. `Remodelling Structural
Realism: Quantum Physics and the Metaphysics of Structure',
forthcoming in \emph{Synthese}.

\vspace{5pt}

\noindent French, S. and Redhead, M. (1988), `Quantum Physics and
the Identity of Indiscernibles', \emph{British Journal for the
Philosophy of Science} 39, 233-246.

\vspace{5pt}

\noindent French, S. and Krause, D. (Forthcoming) \emph{Identity
and Individuality in Quantum Physics}.

\vspace{5pt} \noindent Ginsberg, A. (1981), `Quantum Theory and
the Identity of Indiscernibles Revisited', \emph{Philosophy of
Science} 48, 487-491.

\vspace{5pt} \noindent Greenberg, O.W. (1991), `Interactions of
Particles having Small Violations of Statistics', \emph{Physica}
180, 419-427.

\vspace{5pt} \noindent Greenberg, O.W. and Messiah, A.M.L. (1964),
`Symmetrization Postulate and Its Experimental Foundation',
\emph{Physical Review} 136B, 248-267.

\vspace{5pt}

\noindent Hestenes, D. (1970), `Entropy and Indistinguishability',
\emph{American Journal of Physics} 38, 840-845.

\vspace{5pt} \noindent R. C. Hilborn \& G. M. Tino. (2000),
\emph{Spin-Statistics Connection and Commutation Relations:
Experimental Tests \& Theoretical Implications}. American
Institute of Physics.

\vspace{5pt} \noindent Hoefer, C. (1996), `The Metaphysics of
Spacetime Substantivalism', \emph{Journal of Philosophy} Vol.93,
No.1, 5-27.

\vspace{5pt} \noindent Hilborn, R.C. and Yuca, C.L. (forthcoming),
`Identical Particles in Quantum Mechanics Revisited'.

\vspace{5pt} \noindent Huggett, N. (1994), `What are Quanta, and
Why Does it Matter?', \emph{PSA 1994}, Vol. 2, Philosophy of
Science Association, 69-76.

\vspace{5pt} \noindent Huggett, N. (1997), `Identity, Quantum
Mechanics and Common Sense', \emph{The Monist} 80, 118-130.

\vspace{5pt} \noindent Huggett, N. (1999a), `Atomic Metaphysics',
\emph{J. Phil.}, \textbf{96}, 5-24.

\vspace{5pt} \noindent Huggett, N. (1999b), `On The Significance
of the Permutation Symmetry', \emph{British Journal for the
Philosophy of Science} 50, 325-347.

\vspace{5pt} \noindent Huggett, N. \&  Imbo, T. D.  (2000), `What
is an Elementary Quarticle?', University of Chicago Preprint.

\vspace{5pt} \noindent Imbo, T.D., Shah Imbo, C. and Sudarshan,
E.C.G. (1990), `Identical Particles, Exotic Statistics and Braid
Groups', \emph{Physics Letters} B 234, 103-107.

\vspace{5pt} \noindent Kosso, P. (2000), `The Empirical Status of
Symmetries in Physics', \emph{British Journal for the Philosophy
of Science} 51, 81-98.

\vspace{5pt} \noindent Krause, D. (1992), `On a Quasi-set Theory',
\emph{Notre Dame Journal of Formal Logic}  33, 402--411.

\vspace{5pt} \noindent Ladyman, J. (1998), `What is Structural
Realism?', \emph{Studies in History and Philosophy of Science} 29,
409-424.

\vspace{5pt} \noindent Leinaas, J.M. and Myrheim, J. (1997), `On
the Theory of Identical Particles', \emph{Nuovo Cimento} 37B,
1-23.

\vspace{5pt} \noindent Maidens, A. V. (1993), \emph{The Hole
Argument: Substantivalism and Determinism in General Relativity.}
Ph.D Thesis, University of Cambridge.

\vspace{5pt} \noindent Maidens, A. V. (1998), `Symmetry Groups,
Absolute Objects and Action Principles in General Relativity',
\emph{Stud. Hist. Phil. Mod. Phys.}, Vol. 29, No.2, 245-272.

\vspace{5pt} \noindent Margenau, H. (1944), `The Exclusion
Principle and its Philosophical Importance', \emph{Philosophy of
Science} 11, 187-208.

\vspace{5pt}

\noindent Massimi, M. (2001), `Exclusion Principle and the
Identity of Indiscernibles: A Response to Margenau's Argument',
\emph{British Journal for the Philosophy of Science} 52, 303-330.

\vspace{5pt}
 \noindent   Messiah, A. M. L. (1962), \emph{Quantum Mechanics, Vol. II}.
   Amsterdam: North Holland.

\vspace{5pt} \noindent Miller, A. (1987), `Symmetry and Imagery in
the Physics of Bohr, Einstein and Heisenberg', in \emph{Symmetries
in Physics (1600-1980)}, M.G. Doncel et. al. (eds.). Bellaterra,
300-325.

\vspace{5pt} \noindent Mittelstaedt, P. and Castellani, E. (2000),
`Leibniz's Principle, Physics and the Language of Physics',
\emph{Foundations of Physics} 30, 1585-1604.

\vspace{5pt} \noindent Norton, J. (1988), `The Hole Argument',
\emph{PSA 1988}, Vol. 2, Philosophy of Science Association, 56-64.

\vspace{5pt} \noindent Post, H. (1963), `Individuality and
Physics', \emph{The Listener} 70, 534-537.

\vspace{5pt} \noindent Post, H. (1971), `Correspondence,
Invariance and Heuristics', \emph{Studies in History and
Philosophy of Science} 2, 213-255.

\vspace{5pt} \noindent Quinton, A. (1973), \emph{The Nature of
Things}. Routledge and Kegan Paul.

\vspace{5pt} \noindent   Redhead, M. L. G. (1975), `Symmetry in
Intertheory Relations',
  \emph{Synthese}, 32, 77-112.

\vspace{5pt} \noindent Redhead, M. and Teller, P. (1991),
`Particles, Particle Labels, and Quanta: The Toll of
Unacknowledged Metaphysics', \emph{Foundations of Physics} 21,
43-62.

\vspace{5pt}

 \noindent Redhead, M. and Teller, P. (1992), `Particle Labels and
 the Theory of
 Indistinguishable Particles in Quantum Mechanics',  \emph{British
 Journal for the Philosophy of Science} 43, 201-218.

\vspace{5pt} \noindent Reichenbach, H. (1956), \emph{The Direction
of Time}. University of California Press.

\vspace{5pt} \noindent Russell, B. (1956), `On the Relations of
Universals and Particulars', in \emph{Logic and Knowledge}, R.C.
Marsh (ed.). New York, 105-124.

\vspace{5pt} \noindent Saunders, S. (2002) `Scientific Realism,
Again', forthcoming in \emph{Synthese}.

\vspace{5pt} \noindent Saunders, S. (forthcoming), `Physics and
Leibniz's Law', Oxford University Preprint.

\vspace{5pt} \noindent Stachel, J. (2002), ```The Relations
Between Things' versus `The Things Between Relations': The Deeper
Meaning of the Hole Argument''', in \emph{Reading Natural
Philosophy/ Essays in the History and Philosophy of Science and
Mathematics}, David Malament (ed.). Open Court, Chicago and
LaSalle, Illinois, 231-266.

\vspace{5pt} \noindent Teller, P. (1983), `Quantum Physics, The
Identity of Indiscernibles and Some Unanswered Questions',
\emph{Philosophy of Science} 50, 309-319.

\vspace{5pt} \noindent Teller, P. and Redhead, M. (2001), `Is
Indistinguishability in Quantum Mechanics Conventional?',
\emph{Foundations of Physics} 30, 951-957.

\vspace{5pt} \noindent van Fraassen, B. (1984), `The Problem of
Indistinguishable Particles', in \emph{Science and Reality: Recent
Work in the Philosophy of Science}, J.T. Cushing, C.F. Delaney and
G.M. Gutting (eds.). University of Notre Dame Press, 153-172.

\vspace{5pt} \noindent Van Fraassen, B. (1989), \emph{Laws and
Symmetry}. Oxford University Press.

\vspace{5pt} \noindent van Fraassen, B. (1991), \emph{Quantum
mechanics: An Empiricist View}. Oxford University Press.

\vspace{5pt} \noindent von Meyenn, K. (1987), `Pauli's Belief in
Exact Symmetries', in \emph{Symmetries in Physics (1600-1980)},
M.G. Doncel et. al. (eds.). Bellaterra.

\vspace{5pt} \noindent Weyl, H. (1928), \emph{ The Theory of
Groups and Quantum Mechanics}. Methuen and Co.; English trans.2nd
ed..

\vspace{5pt} \noindent H, Weyl. (1952), \emph{Symmetry}. Princeton
University Press.

\end{document}